\documentclass[12pt,twoside]{article}
\usepackage{fleqn,espcrc1}
\usepackage{psfig}

\newcommand{\AmS}{{\protect\the\textfont2
  A\kern-.1667em\lower.5ex\hbox{M}\kern-.125emS}}

\title{Stellar and nuclear-physics constraints on two r-process components
in the early Galaxy}

\author{
B.~Pfeiffer\address[KCHM]{Institut f\"ur Kernchemie, 
Universit\"at Mainz, D-55128 Mainz, Germany}\address[MP]{Max-Planck-Institut 
f\"ur Chemie, Department of Cosmochemistry,
D-55128 Mainz, Germany},
U.~Ott\addressmark[MP], K.-L.~Kratz\addressmark[KCHM]
}

\begin{document}

\maketitle


Recent astrophysical results indicate the existence of (at least) two types of 
the rapid neutron-capture nucleosynthesis (r-process).
The evidence is based on a variety of observations in different fields: \\
(a) The study of extinct radionuclides present in the early Solar System
\cite{was96,qian}.\\
(b) Isotope abundance anomalies observed in presolar diamonds 
\cite{ott96,rich98}. \\
(c) The strongest -- because less model-dependent -- indication for more than 
one type of r-process, however, may come from the observation of heavy
neutron-capture element abundances in very metal-poor halo stars 
\cite{wes99,burris,jaj,sned00} as well as in the globular cluster M15 
\cite{m15}. 

On the one hand, metallicity-scaled abundances 
of elements in the Pt peak and down to Ba (Z=56) in all halo stars so far
investigated are in remarkable agreement 
with the solar r-process pattern (N$_{r,\odot}$), while on the other hand the 
abundances of 
``low-Z'' neutron-capture elements ($_{39}$Y to $_{48}$Cd) in CS 22892-052
are lower than solar \cite{sned00}.
An interesting feature of the abundances of these elements 
is their pronounced
odd-even-Z staggering, which reflects nuclear-structure properties of the
progenitor isotopes involved. All odd-Z elements from $_{39}$Y to
$_{47}$Ag are clearly under-abundant compared to the
solar pattern, whereas the even-Z elements
($_{40}$Zr -- $_{48}$Cd) are closer to solar (see Fig.~1). 

\begin{figure}
\centerline{\psfig{file=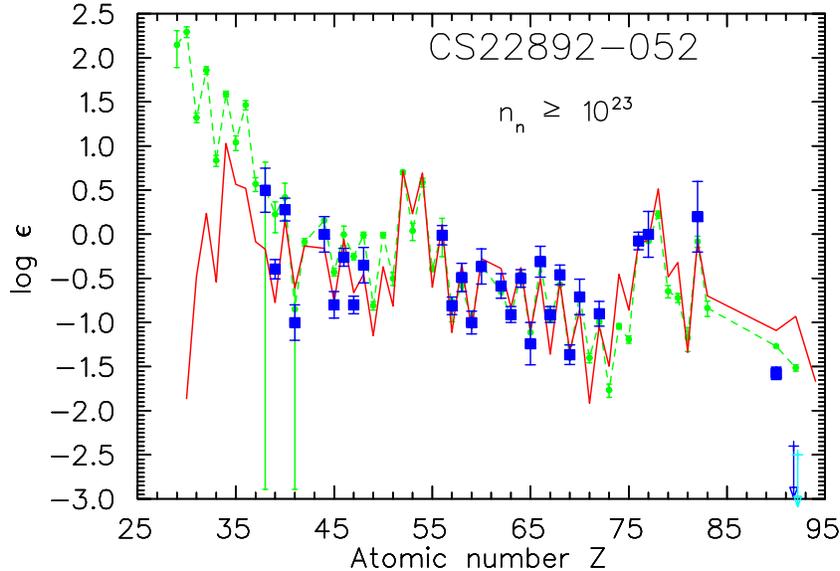,angle=90,width=11cm}}
\caption{
Comparison between observed (filled squares) and calculated (solid line) 
elemental r-abundances from the ultra-metal-poor halo star CS 22892-052.
The abundance distribution from Z$\simeq$40 to $_{90}$Th is denoted as 
the {\it ``main''} r-process in the text. The scaled solar-system 
distribution is shown as dashed curve with filled circles. The N$_{r,\odot}$ -
N$_{r,main}$ ``residuals'' at ``low-Z'' require contributions from a second
({\it ``weak''}) r-process; see Fig.~2.}
\label{fig1}
\end{figure}

Taking advantage of our site-independent waiting-point approach to fit the 
N$_{r,\odot}$ pattern, 
we now can test under which stellar conditions the possible two r-processes,
presumably separated by the A$\simeq$130 N$_{r,\odot}$ peak, have to run.
When assuming that the abundances 
are a living record of the first (few) generation(s) of Galactic
nucleosynthesis \cite{cow99}, the observed pattern beyond
Z${\simeq}$40 up to $_{90}$Th should most likely be produced by
only one (or a few) r-process event(s) in a unique stellar site, e.g. 
supernovae of type II (SNII). This scenario
(the {\it ``main''} r-process) then produces the ''low-Z'' elements
under-abundant compared to solar, and reaches the full solar values 
presumably around $_{52}$Te. For CS22892-052 both, the general trend as well 
as the detailed structure of the ''low-Z'' abundances
(40${\leq}$Z${\leq}$48) are nicely reproduced in our fit with the
ETFSI-Q atomic masses (see Fig.~1). At the same time, the good overall 
reproduction of
the ''high-Z'' elements (beyond $_{56}$Ba) is maintained \cite{jrnc}. 
Starting our calculations from an Fe-group seed would require neutron 
densities of n$_n$$\ge$10$^{23}$ cm$^{-3}$ at freeze-out (T$_9$=1.35). 
It should be mentioned in this context, that our approach  would 
imply a roughly constant abundance ratio between the ``low-Z'' and ``high-Z''
elements. This has recently been confirmed in the case of HD115444, where our
prediction for Ag agrees with the observation \cite{wes99}.

Consequently, the abundance {\it ''residuals''} 
(N$_{r,\odot}$--N$_{r,main}$=N$_{r,resid}$) at low Z will require a separate
{\it ``weak''} r-process component of yet unknown stellar
site.  When assuming seed compositions from $_{14}$Si to $_{24}$Cr or $_{28}$Ni
in solar-system fractions, our calculations can reproduce the N$_{r,weak}$  
pattern in CS22892-052
with neutron densities of n$_n$$\leq$10$^{20}$ cm$^{-3}$ and process 
durations $\tau$$\simeq$500--1000 ms (see Fig.~2). These stellar conditions  
might be provided in explosive shell-burning scenarios (see, e.g. \cite{fr,tr}).
\begin{figure}
\centerline{\psfig{file=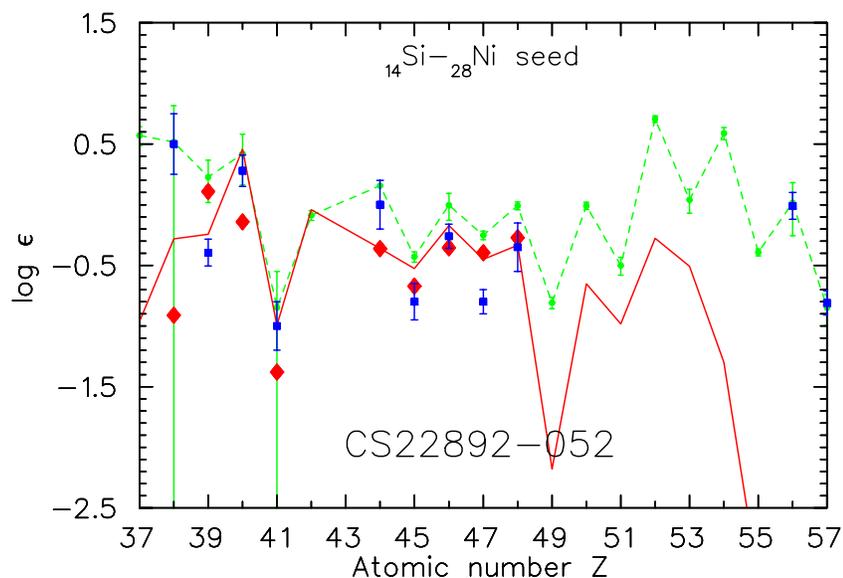,angle=90,width=11.cm}}
\caption{
Comparison between abundance {\it ``residuals''} 
(N$_{r,\odot}$--N$_{halo}$=N$_{r,resid}$; filled diamonds) and calculated 
(full curve) elemental r-abundances from the ultra-metal-poor halo star 
CS 22892-052.
This abundance distribution for ``low-Z'' elements is denoted as 
the ``weak'' r-process in the text. The scaled solar-system 
distribution is shown as dashed curve with filled circles, the observed 
halo-values are displayed as filled squares.}
\label{fig2}
\end{figure}

The {\it ``weak''} component as identified here must be of secondary origin,
as is clearly shown by its absence in the old metal-poor halo stars 
\cite{sned00}. In contrast, the presence there of the main component with a
pattern virtually identical to that of the solar system r-process in the 
mass range above A$\simeq$130--140 attests to its primary and robust nature.
Another outcome of our calculations is that the {\it ``weak''} component as
devised to produce the low-mass r-process nuclides in solar-system proportions
that are ``missing'' from the {\it ``main''} component does not make a 
significant contribution to the A$\simeq$130 abundance peak, in agreement with
calculations from Truran and Cowan \cite{tr}.  Our result thus does not 
support the
conclusion of Qian et al. \cite{qian} of seperate r-process sources being 
responsible for the observed abundance level in the early solar system of
extinct radionuclides $^{129}$I and $^{182}$Hf. In this context, we note that
in all models \cite{qian,cow99,jrnc} the actinides are co-produced with the 
nuclides in the Hf range, but that the observed limit on the abundance in the 
early solar system of $^{247}$Cm ($^{247}$Cm/$^{235}$U$<$4$\times$10$^{-3}$;
\cite{chen}) is barely compatible with expectations based on the same
approach as used for $^{182}$Hf. An improved measurement of this abundance
ratio may be an important step to address the question whether or not for 
$^{182}$Hf a special process \cite{meyer} is required.

\end{document}